# Polynomial-Time, Semantically-Secure Encryption Achieving the Secrecy Capacity

Mihir Bellare[1] Stefano Tessaro[2]

September 2011

**Abstract**

In the wiretap channel setting, one aims to get information-theoretic privacy of communicated data based only on the assumption that the channel from sender to adversary is noisier than the one from sender to receiver. The secrecy capacity is the optimal (highest possible) rate of a secure scheme, and the existence of schemes achieving it has been shown. For thirty years the ultimate and unreached goal has been to achieve this optimal rate with a scheme that is polynomial-time. (This means both encryption and decryption are proven polynomial time algorithms.) This paper finally delivers such a scheme. In fact it does more. Our scheme not only meets the classical notion of security from the wiretap literature, called MIS-R (mutual information security for random messages) but achieves the strictly stronger notion of semantic security, thus delivering more in terms of security without loss of rate.

[1] Department of Computer Science & Engineering, University of California San Diego, 9500 Gilman Drive, La Jolla, California 92093, USA. Email: `mihir@cs.ucsd.edu`. URL: `http://www.cs.ucsd.edu/users/mihir`. Supported in part by NSF grants CNS-0904380 and CCF-0915675.

[2] Department of Computer Science & Engineering, University of California San Diego, 9500 Gilman Drive, La Jolla, California 92093, USA. Email: `stessaro@cs.ucsd.edu`. URL: `http://www.cs.ucsd.edu/users/stessaro`. Supported in part by Calit2 and NSF grant CNS-0716790.

# Contents

# 1 Introduction

Introduced by Wyner, Csiszár and Körner in the late seventies [34, 11], the wiretap channel is a setting where one aims to obtain information-theoretic security (privacy) of communicated data under the sole assumption that the channel from sender to adversary is "noisier" than the channel from sender to receiver. Researchers have shown that there is a maximum possible rate (ratio of message length to ciphertext length) for a secure scheme, called the optimal rate, and they have shown, through the probabilistic method, that there exist secure schemes with this rate. But these results are non-constructive. A question of great interest in this area is whether there is an explicit, secure scheme that is polynomial-time. (Meaning, there are polynomial-time algorithms for both encryption and decryption.) But this has remained open for 30 years. In this paper we finally resolve this question by providing such a scheme.

However, we do even more. Our scheme achieves not only the classical notion of security from the wiretap literature but the stronger notion of semantic (equivalently, distinguishing) security of [3]. Furthermore our scheme is simple, efficient and modular. Unlike schemes from the I&C approaches, it makes only blackbox (meaning non-intrusive) use of error-correcting codes. Our scheme is obtained by combining methods from cryptography and coding theory. Let us now look at all this in some more detail.

The wiretap model. The setting is depicted in Figure 1. The sender applies to her message $\mathsf{M}$ a randomized encryption function $\mathcal{E}\colon \{0,1\}^m \to \{0,1\}^c$ to get what we call the *sender-ciphertext* $\mathsf{X} \leftarrow\!\!\$\, \mathcal{E}(\mathsf{M})$.[1] This is transmitted to the receiver over the receiver channel $\mathsf{ChR}$ so that the latter gets a *receiver ciphertext* $\mathsf{Y} \leftarrow\!\!\$\, \mathsf{ChR}(\mathsf{X})$ which it decrypts via algorithm $\mathcal{D}$ to recover the message. The adversary's wiretap is modeled as another channel $\mathsf{ChA}$ and it accordingly gets an *adversary ciphertext* $\mathsf{Z} \leftarrow\!\!\$\, \mathsf{ChA}(\mathsf{X})$ from which it tries to glean whatever it can about the message.

A *channel* is a randomized function specified by a transition probability matrix $W$ where $W[x, y]$ is the probability that input $x$ results in output $y$. Here $x, y$ are strings. Thus, for example, we regard the Binary Symmetric Channel $\mathsf{BSC}_p$ with crossover probability $p \le 1/2$ as taking a binary string $x$ of any length and returning the string $y$ of the same length formed by flipping each bit of $x$ independently with probability $p$. For concreteness and simplicity of exposition we will often phrase discussions in the setting where $\mathsf{ChR}, \mathsf{ChA}$ are BSCs with crossover probabilities $p_R, p_A \le 1/2$ respectively, but our results apply in much greater generality. In this case the assumption that $\mathsf{ChA}$ is "noisier" than $\mathsf{ChR}$ corresponds to the assumption that $p_R < p_A$. This is the only assumption made: the adversary is computationally unbounded, and the scheme is keyless, meaning sender and receiver are not assumed to a priori share any information not known to the adversary.

Requirements. The two requirements are *decryptability*, also called *decodability*, and *security*. The first asks that the scheme provide error-correction over the receiver channel, namely $\lim_{m\to\infty} \Pr[\mathcal{D}(\mathsf{ChR}(\mathcal{E}(\mathsf{M}))) \neq \mathsf{M}] = 0$. Security may be measured in various ways. A security *metric* xs associates to encryption function $\mathcal{E}\colon \{0,1\}^m \to \{0,1\}^c$ and adversary channel $\mathsf{ChA}$ a number $\mathbf{Adv}^{\mathrm{xs}}(\mathcal{E}; \mathsf{ChA})$ that measures the maximum "advantage" of an adversary in breaking the scheme under metric xs, and we say that $\mathcal{E}$ provides XS-security relative to $\mathsf{ChA}$ if $\lim_{k\to\infty} \mathbf{Adv}^{\mathrm{xs}}(\mathcal{E}; \mathsf{ChA}) = 0$. The central metric of the wiretap literature is mis-r (mutual-information security for random messages), defined via $\mathbf{Adv}^{\mathrm{mis\text{-}r}}(\mathcal{E}; \mathsf{ChA}) = \mathbf{I}(\mathsf{M}; \mathsf{ChA}(\mathcal{E}(\mathsf{M})))$ where $\mathsf{M}$ is uniformly distributed over $\{0,1\}^m$ and $\mathbf{I}$ is the mutual information. It was introduced by [26, 27] and strengthens the original metric of [34]. The name is from [3].

Taking a cryptographic perspective, the latter point out that mis-r is weak because messages are assumed to be random. They introduce a semantic security (ss) metric following [15] that, roughly, asks that given the adversary ciphertext, the adversary cannot compute any function of the message with probability better than she could have without the adversary ciphertext. They show that this is equivalent to a strengthening of mis-r that they call mis security and also to a simpler and more convenient metric called distinguishing security (ds), also adapted from [15], where the advantage is defined via

$$\mathbf{Adv}^{\mathrm{ds}}(\mathcal{E}; \mathsf{ChA}) \;=\; \max_{\mathcal{A}, M_0, M_1} \; 2\Pr[\mathcal{A}(M_0, M_1, \mathsf{ChA}(\mathcal{E}(M_{\mathsf{b}}))) = \mathsf{b}] - 1$$

[1] The notation $y \leftarrow\!\!\$\, A(x)$ means that we run randomized function $A$ on input $x$ and denote the output by $y$.

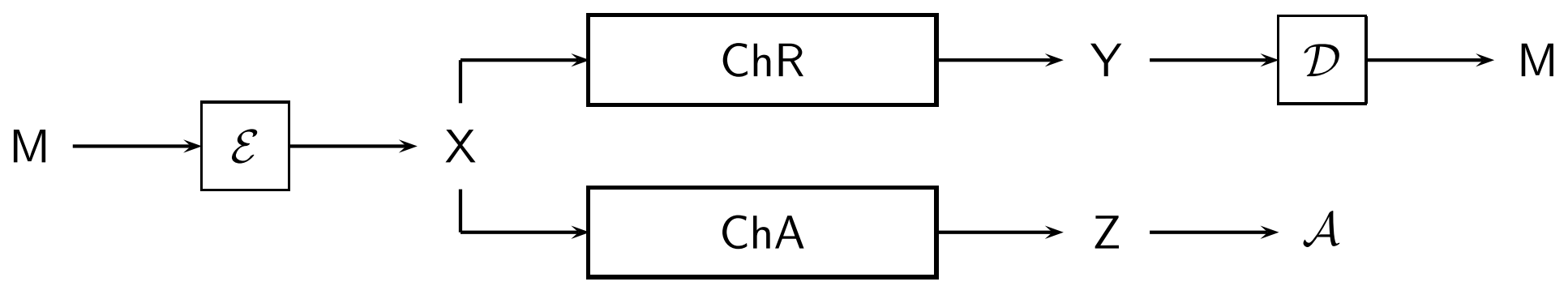

Figure 1: **Wiretap channel model.** See text for explanations.

where challenge bit b is uniformly distributed over $\{0,1\}$ and the maximum is over all $m$-bit messages $M_0, M_1$ and all adversaries $\mathcal{A}$. Since MIS, SS and DS are shown equivalent in [3] one can work with any of them, and our choice is DS.

Practical interest in the wiretap setting is escalating [35, 4], and applications need DS-security rather than MIS-R security. Thus DS-security is the most desirable target.

Previous work. In the Information and Coding (I&C) community, the wiretap setting has a literature encompassing hundreds of papers. (See the survey [23] or the book [4].) The focus has been to show the existence of MIS-R-secure schemes with optimal rate. (The schemes are not required to be explicit, let alone polynomial time.) This optimal rate is called the secrecy capacity. In the case of BSCs, it equals the difference $(1 - h_2(p_R)) - (1 - h_2(p_A)) = h_2(p_A) - h_2(p_R)$ in capacities of the receiver and adversary channels, where $h_2(p) = -p\lg(p) - (1-p)\lg(1-p)$ is the binary entropy. Non-constructive proofs of the existence of MIS-R-secure schemes with this optimal rate were given in [34, 11, 5]. A lot of work has followed aiming to establish similar results for other channels.

Mahdavifar and Vardy [24, 25] provide an explicit MIS-R-secure scheme with optimal rate, but they give no proof that decoding is possible for their scheme, even in principle let alone in polynomial time. The central open question in the wiretap channel community was whether there is a polynomial time (this means both encryption and decryption are polynomial time) MIS-R secure scheme with optimal rate.

DS-security has upped the ante. The first question here is to determine this optimal rate. Since DS-security is stronger than MIS-R security, the optimal rate could in principle be smaller but (perhaps surprisingly), for a broad class of channels, it isn't. That is, the optimal rate is the same for DS and MIS-R security for a broad class of channels including symmetric channels. This follows by applying a result of [3], which shows that MIS-R implies MIS for certain types of schemes and channels, to the scheme of [24, 25].

Polynomial-time DS-secure schemes were presented in [3] but their rate is not optimal. In summary, the most desirable goal here is to not only solve the long-standing open question from the wiretap community by giving a polynomial-time MIS-R-secure scheme with optimal rate but go further and give a polynomial-time DS-secure scheme with optimal rate.

Our scheme. This paper resolves the above open problem, providing the first polynomial-time scheme that achieves DS (and hence MIS-R) security with optimal rate, meaning rate equal to the secrecy capacity.

The scheme of [24, 25] is based on polar codes [2]. Our approach is modular and is able to use any ECC, so that we do not rely on the structure of specific ECCs.

One might hope to build a scheme for the case where the receiver channel is noiseless and then add error-correction to meet the decoding condition with a noisy receiver channel. This does not work because the error-correction helps the adversary by reducing the noise over the adversary channel. The need to to couple security and decoding considerations in the design is one source of challenges.

Our scheme is based on three main ideas: the use of *invertible* extractors; analysis via smooth min-entropy; and an adaption of the result of [3] saying that for certain types of schemes, DS-security on random messages implies DS-security on all messages. Section 5 overviews the technical approach, describes the components and scheme in detail and proves DS security.

We are stating asymptotic results for simplicity. Our proof will show a quantitative bound on adversary ds-advantage that decays exponentially with the security parameter. The scheme is also fairly simple and efficient. Finally the claims (proven DS-security and decoding with optimal rate) hold not only for BSCs but for a wide range of receiver and adversary channels.

A CONCRETE INSTANTIATION. As a consequence of our general paradigm, we prove, for example, that the following simple scheme achieves secrecy capacity for the setting where $\mathsf{ChR}$ and $\mathsf{ChA}$ are BSC's with respective crossover probabilities $p_R < p_A \leq 1/2$. Let $\mathsf{E}\colon \{0,1\}^k \to \{0,1\}^n$ be an error-correcting code which is efficiently decodable for the BSC with crossover probability $p_R$, and such that $k \approx (1-h_2(p_R))\cdot n$ (such ECCs can be built e.g. from polar codes [2] or from concatenated codes [14]). Our encryption function $\mathcal{E}$ takes as input an $m$-bit message $M$, where $m = b \cdot t$, $b \approx (h_2(p_A) - h_2(p_R)) \cdot n$, and $t$ is a parameter of the scheme. It first chooses uniformly at random a $k$-bit string $A \neq 0^k$ as well as $t$ $(k-b)$-bit strings $R[1], \ldots, R[t]$. It then splits $M$ into $t$ $b$-bit blocks $M[1], \ldots, M[t]$, and outputs

$$\mathcal{E}(M) = \mathsf{E}(A) \parallel \mathsf{E}(A \odot (M[1] \parallel R[1])) \parallel \cdots \parallel \mathsf{E}(A \odot (M[t] \parallel R[t])) ,$$

where $\odot$ is multiplication of $k$-bit strings interpreted as elements of the extension field $\mathrm{GF}(2^k)$.

RELATED WORK. Appendix A surveys related to wiretap security.

## 2 Preliminaries

BASIC NOTATION AND DEFINITIONS. If $s$ is a binary string then $s[i]$ denotes its $i$-th bit and $|s|$ denotes its length. If $S$ is a set then $|S|$ denotes its size. If $x$ is a real number then $|x|$ denotes its absolute value. If $s_1, \ldots, s_l$ are strings then $s_1 \| \cdots \| s_l$ denotes their concatenation. If $s$ is a string and $n$ a non-negative integer then $s^n$ denotes the concatenation of $n$ copies of $s$.

A probability distribution is a function $P$ that associates to each $x$ a probability $P(x) \in [0,1]$. The support $\mathrm{SUPP}(P)$ is the set of all $x$ such that $P(x) > 0$. All probability distributions in this paper are discrete. Associate to random variable $\mathsf{X}$ and event $E$ the probability distributions $P_X, P_{X|E}$ defined for all $x$ by $P_{\mathsf{X}}(x) = \Pr[\mathsf{X} = x]$ and $P_{\mathsf{X}|E}(x) = \Pr[\mathsf{X} = x \mid E]$. We denote by $\lg(\cdot)$ the logarithm in base two, and by $\ln(\cdot)$ the natural logarithm. We adopt standard conventions such as $0 \lg 0 = 0 \lg \infty = 0$ and $\Pr[E_1|E_2] = 0$ when $\Pr[E_2] = 0$. The function $h\colon [0,1] \to [0,1]$ is defined by $h(x) = -x \lg x$. The (Shannon) entropy of probability distribution $P$ is defined by $\mathbf{H}(P) = \sum_x h(P(x))$ and the statistical difference between probability distributions $P, Q$ is defined by $\mathbf{SD}(P; Q) = 0.5 \cdot \sum_x |P(x) - Q(x)|$. If $\mathsf{X}, \mathsf{Y}$ are random variables the (Shannon) entropy is defined by $\mathbf{H}(\mathsf{X}) = \mathbf{H}(P_{\mathsf{X}}) = \sum_x h(P_{\mathsf{X}}(x))$. The conditional entropy is defined via $\mathbf{H}(\mathsf{X} \mid \mathsf{Y} = y) = \sum_x h(P_{\mathsf{X}|\mathsf{Y}=y}(x))$ and $\mathbf{H}(\mathsf{X} \mid \mathsf{Y}) = \sum_y P_{\mathsf{Y}}(y) \cdot \mathbf{H}(\mathsf{X} \mid \mathsf{Y} = y)$. The statistical or variational distance between random variables $\mathsf{X}_1, \mathsf{X}_2$ is $\mathbf{SD}(\mathsf{X}_1; \mathsf{X}_2) = \mathbf{SD}(P_{\mathsf{X}_1}; P_{\mathsf{X}_2}) = 0.5 \cdot \sum_x |\Pr[\mathsf{X}_1 = x] - \Pr[\mathsf{X}_2 = x]|$. The min-entropy of random variable $\mathsf{X}$ is $\mathbf{H}_\infty(\mathsf{X}) = \max_x \Pr[\mathsf{X} = x]$ and if $\mathsf{Z}$ is also a random variable the conditional min-entropy is $\mathbf{H}_\infty(\mathsf{X}|\mathsf{Z}) = \sum_z \Pr[\mathsf{Z} = z] \max_x \Pr[\mathsf{X} = x|\mathsf{Z} = z]$.

TRANSFORMS, CHANNELS AND ALGORITHMS. We say that $T$ is a transform with domain $D$ and range $R$, written $T\colon D \to R$, if $T(x)$ is a random variable over $R$ for every $x \in D$. Thus, $T$ is fully specified by a sequence $P = \{P_x\}_{x \in D}$ of probability distributions over $R$, where $P_x(y) = \Pr[T(x) = y]$ for all $x \in D$ and $y \in R$. We call $P$ the distribution associated to $T$. This distribution can be specified by a $|D|$ by $|R|$ transition probability matrix $W$ defined by $W[x, y] = P_x(y)$. A (randomized) algorithm is also a transform. Finally, an adversary too is a transform.

CHANNELS. A channel is, again, just a transform. In more conventional communications terminology, a channel $\mathsf{Ch}\colon D \to R$ has input alphabet $D$ and output alphabet $R$.

If $\mathsf{B}\colon D \to Z$ is a channel and $c \geq 1$ is an integer we define the channel $\mathsf{B}^c\colon \{0,1\}^c \to Z^c$ by $\mathsf{B}^c(X) = \mathsf{B}(X[1]) \| \cdots \| \mathsf{B}(X[c])$ for all $X = X[1] \ldots X[c] \in \{0,1\}^c$. The applications of $\mathsf{B}$ are all independent, meaning that if $W$ is the transition probability matrix of $\mathsf{B}$ then the transition probability matrix $W_c$ of $\mathsf{B}^c$ is defined by $W[X, Y] = W[X[1], Y[1]] \cdot \ldots \cdot W[X[c], Y[c]]$ for all $X = X[1] \ldots X[c] \in \{0,1\}^c$ and $Y = Y[1] \ldots Y[c] \in Z^c$. We say that a channel $\mathsf{Ch}$ is binary if it equals $\mathsf{B}^c$ for some channel $\mathsf{B}$ and some $c$, in which case we refer to $\mathsf{B}$ as the base (binary) channel and $\mathsf{Ch}$ as the channel induced by $\mathsf{B}$.

By $\mathsf{BSC}_p\colon \{0,1\} \to \{0,1\}$ we denote the binary symmetric channel with crossover probability $p$ $(0 \leq p \leq 1/2)$. Its transition probability matrix $W$ has $W[x, y] = p$ if $x \neq y$ and $1 - p$ otherwise for all $x, y \in \{0,1\}$. The induced channel $\mathsf{BSC}_p^c$ flips each input bit independently with probability $p$.

The receiver and adversary channels of the wiretap setting will have domain $\{0,1\}^c$, where $c$ is the length of the sender ciphertext, and range $\{0,1\}^d$, where the output length $d$ may differ between the two channels. Such channels may be binary, which is the most natural example, but our equivalences between security notions hold for all channels, even ones that are not binary.

If $\mathsf{Ch1}\colon \{0,1\}^{c_1} \to \{0,1\}^{d_1}$ and $\mathsf{Ch2}\colon \{0,1\}^{c_2} \to \{0,1\}^{d_2}$ are channels then $\mathsf{Ch1}\|\mathsf{Ch2}$ denotes the channel $\mathsf{Ch}\colon \{0,1\}^{c_1+c_2} \to \{0,1\}^{d_1+d_2}$ defined by $\mathsf{Ch}(x_1\|x_2) = \mathsf{Ch1}(x_1)\|\mathsf{Ch2}(x_2)$ for all $x_1 \in \{0,1\}^{c_1}$ and $x_2 \in \{0,1\}^{c_2}$.

Finally, we say that a channel $\mathsf{Ch}\colon D \to R$ with transition matrix $W$ is *symmetric* if the there exists a partition of the range as $R = R_1 \cup \cdots \cup R_n$ such that for all $i$ the sub-matrix $W[\cdot, R_i]$ induced by the rows in $R_i$ is strongly symmetric, i.e., all rows are permutations of each other, and all columns are permutations of each other.

# 3 Encryption and Semantic Security

Our formalization of encryption functions and schemes, as well as their security, follows the approach of [3]. We briefly review the main tools, and refer the reader to [3] for further details.

Encryption functions and schemes. An *encryption function* is a transform $\mathcal{E}\colon \{0,1\}^m \to \{0,1\}^c$ where $m$ is the message length and $c$ is the sender ciphertext length. The *rate* of $\mathcal{E}$ is $\mathbf{Rate}(\mathcal{E}) = m/c$. If $\mathsf{ChR}\colon \{0,1\}^c \to \{0,1\}^d$ is a receiver channel then a *decryption function* for $\mathcal{E}$ over $\mathsf{ChR}$ is a transform $\mathcal{D}\colon \{0,1\}^d \to \{0,1\}^m$ whose decryption error $\mathbf{DE}(\mathcal{E};\mathcal{D};\mathsf{ChR})$ is defined as the maximum, over all $M \in \{0,1\}^m$, of $\Pr[\mathcal{D}(\mathsf{ChR}(\mathcal{E}(M))) \neq M]$.

An *encryption scheme* $\overline{\mathcal{E}} = \{\mathcal{E}_k\}_{k\in\mathbb{N}}$ is a family of encryption functions where $\mathcal{E}_k\colon \{0,1\}^{m(k)} \to \{0,1\}^{c(k)}$ for functions $m, c\colon \mathbb{N} \to \mathbb{N}$ called the message length and sender ciphertext lengths of the scheme. Suppose $\overline{\mathsf{ChR}} = \{\mathsf{ChR}_k\}_{k\in\mathbb{N}}$ is a family of receiver channels where $\mathsf{ChR}_k\colon \{0,1\}^{c(k)} \to \{0,1\}^{d(k)}$. Then a *decryption scheme* for $\overline{\mathcal{E}}$ over $\overline{\mathsf{ChR}}$ is a family $\overline{\mathcal{D}} = \{\mathcal{D}_k\}_{k\in\mathbb{N}}$ where $\mathcal{D}_k\colon \{0,1\}^{d(k)} \to \{0,1\}^{m(k)}$ is a decryption function for $\mathcal{E}_k$ over $\mathsf{ChR}_k$. The *decoding* requirement, also called the *decryption* requirement, is that $\lim_{k\to\infty} \mathbf{DE}(\mathcal{E}_k;\mathcal{D}_k;\mathsf{ChR}_k) = 0$. The *rate* of $\overline{\mathcal{E}}$ is $\mathbf{Rate}(\overline{\mathcal{E}}) = \lim_{k\to\infty} \mathbf{Rate}(\mathcal{E}_k)$.

We say that a family $\{\mathcal{S}_k\}_{k\in\mathbb{N}}$ (eg. an encryption or decryption scheme) is polynomial-time computable if there is a polynomial time computable function which on input $1^k$ (the unary representation of $k$) and $x$ returns $\mathcal{S}_k(x)$. Our constructs will provide polynomial-time computable encryption and decryption schemes.

Semantic security. Let $\mathcal{E}\colon \{0,1\}^m \to \{0,1\}^c$ be an encryption function and let $\mathsf{ChA}\colon \{0,1\}^c \to \{0,1\}^d$ be an adversary channel. Security depends only on these, not on the receiver channel. Following [3], in this paper we will target semantic security (ss) and distinguishing security (ds). We refer the reader to [3] for an in depth study of these notions, and their relation to entropy-based security metrics.

Concretely, the ss advantage is defined as

$$\mathbf{Adv}^{\mathrm{ss}}(\mathcal{E};\mathsf{ChA}) \;=\; \max_{f,\mathsf{M}} \left( \max_{\mathcal{A}} \Pr[\mathcal{A}(\mathsf{ChA}(\mathcal{E}(\mathsf{M}))) = f(\mathsf{M})] - \max_{\mathcal{S}} \Pr[\mathcal{S}(m) = f(\mathsf{M})] \right) , \tag{1}$$

where $f$ is a transform with domain $\{0,1\}^m$ that represents partial information about the message. Moreover, the distinguishing advantage is

$$\begin{aligned} \mathbf{Adv}^{\mathrm{ds}}(\mathcal{E};\mathsf{ChA}) &= \max_{\mathcal{A},M_0,M_1} 2\Pr[\mathcal{A}(M_0,M_1,\mathsf{ChA}(\mathcal{E}(M_\mathsf{b}))) = \mathsf{b}] - 1 & (2)\\ &= \max_{M_0,M_1} \mathbf{SD}(\mathsf{ChA}(\mathcal{E}(M_0));\mathsf{ChA}(\mathcal{E}(M_1))) , & (3) \end{aligned}$$

where $\Pr[\mathcal{A}(M_0,M_1,\mathsf{ChA}(\mathcal{E}(M_\mathsf{b}))) = \mathsf{b}]$ is the probability that adversary $\mathcal{A}$, given $m$-bit messages $M_0, M_1$ and an adversary ciphertext emanating from $M_\mathsf{b}$, correctly identifies the random challenge bit $\mathsf{b}$. We note that this advantage is equal to the statistical distance between the random variables $\mathsf{ChA}(\mathcal{E}(M_0))$ and $\mathsf{ChA}(\mathcal{E}(M_1))$.

We say that the encryption scheme $\overline{\mathcal{E}} = \{\mathcal{E}_k\}_{k\in\mathbb{N}}$ is SS-secure relative to $\overline{\mathsf{ChA}} = \{\mathsf{ChA}_k\}_{k\in\mathbb{N}}$ if $\lim_{k\to\infty} \mathbf{Adv}^{\mathrm{ss}}(\mathcal{E}_k; \mathsf{ChA}_k) = 0$. This does not mandate any particular rate at which the advantage should vanish, but in our constructions this rate is exponentially vanishing with $k$. Similarly, the scheme is DS-secure if $\lim_{k\to\infty} \mathbf{Adv}^{\mathrm{ds}}(\mathcal{E}_k; \mathsf{ChA}_k) = 0$. The following theorem, proved in [3], establishes the equivalence of SS- and DS-security.

**Theorem 3.1** **[DS $\leftrightarrow$ SS]** *Let* $\mathcal{E}\colon \{0,1\}^m \to \{0,1\}^c$ *be an encryption algorithm and* $\mathsf{ChA}$ *an adversary channel. Then* $\mathbf{Adv}^{\mathrm{ss}}(\mathcal{E}; \mathsf{ChA}) \le \mathbf{Adv}^{\mathrm{ds}}(\mathcal{E}; \mathsf{ChA}) \le 2 \cdot \mathbf{Adv}^{\mathrm{ss}}(\mathcal{E}; \mathsf{ChA})$. ▮

# 4 Seeded Encryption

We introduce an extension of the standard wiretap setting where Alice, Bob, and Eve have access to a common random string $S$, called the *seed*, chosen honestly and randomly. This setting is interesting in its own right: One can think of the seed as being chosen once and for all when deploying an encryption scheme. More importantly, however, seeded encryption can be seen as an intermediate step towards building a regular (unseeded) encryption scheme, as we explain below.

SEEDED ENCRYPTION. A *seeded encryption function* is a transform $\mathcal{SE}\colon \mathrm{SDS} \times \{0,1\}^b \to \{0,1\}^n$ that takes a seed $S \in \mathrm{SDS}$ and message $M \in \{0,1\}^b$ to return a sender ciphertext denoted $\mathcal{SE}(S, M)$ or $\mathcal{SE}_S(M)$, so that each seed $S$ defines an encryption function $\mathcal{SE}_S\colon \{0,1\}^b \to \{0,1\}^n$. Given a channel $\mathsf{ChR} : \{0,1\}^n \to \{0,1\}^\ell$, a *seeded decryption function* $\mathcal{SD}$ *for* $\mathcal{SE}$ *over* $\mathsf{ChR}$ is a a transform $\mathcal{SD} : \mathrm{SDS} \times \{0,1\}^\ell \to \{0,1\}^b$. The decryption error $\mathbf{DE}(\mathcal{SE}; \mathcal{SD}; \mathsf{ChR})$ of $\mathcal{SE}$, $\mathcal{SD}$, and $\mathsf{ChR}$ is defined as

$$\mathbf{DE}(\mathcal{SE}; \mathcal{SD}; \mathsf{ChR}) = \mathbf{E}\left[\max_{M\in\{0,1\}^b} \Pr\left[\mathcal{SD}(S, \mathsf{ChR}(\mathcal{SE}(S, M))) \neq M\right]\right],$$

where the expectation is taken over the random choice of $S$. A *seeded encryption scheme* is a family $\overline{\mathcal{SE}} = \{\mathcal{SE}_k\}_{k\in\mathbb{N}}$. The rate $\mathbf{Rate}(\mathcal{SE})$ of a seeded encryption function $\mathcal{SE}$ is defined as $b/c$, meaning the seed is ignored, and, accordingly, we let $\mathbf{Rate}(\overline{\mathcal{SE}}) = \lim_{k\to\infty} \mathbf{Rate}(\mathcal{SE}_k)$.

DISTINGUISHING SECURITY FOR SEEDED ENCRYPTION. We extend distinguishing security to the setting of seeded encryption. It is defined via a game where the adversary is first given the seed $S \leftarrow\$\ \mathrm{SDS}$. It then outputs two messages $M_0, M_1$, and is subsequently given the encryption $\mathsf{ChA}(\mathcal{SE}_S(M_{\mathsf{b}}))$ for a random bit $\mathsf{b}$. Finally, it outputs a bit $b'$, and wins the game if $\mathsf{b} = b'$. As the adversary can choose the best pair of message $M_0, M_1$ for each choice of the seed, the optimal strategy guesses the bit $\mathsf{b}$ with probability $(1 + \mathbf{Adv}^{\mathrm{ds}}(\mathcal{SE}_S; \mathsf{ChA}))/2$ conditioned on a particular choice of the seed $S$. Therefore, the optimal adversary guesses $\mathsf{b}$ with probability

$$\mathbf{E}\left[\frac{1 + \mathbf{Adv}^{\mathrm{ds}}(\mathcal{SE}_S; \mathsf{ChA}))}{2}\right] = \frac{1}{2} + \frac{\mathbf{E}\left[\,\mathbf{Adv}^{\mathrm{ds}}(\mathcal{SE}_S; \mathsf{ChA})\,\right]}{2},$$

where the expectations are over $S$ drawn at random from $\mathrm{SDS}$, and equality follows from linearity of expectations. Consequently, we define the ds advantage as

$$\mathbf{Adv}^{\mathrm{ds}}(\mathcal{SE}; \mathsf{ChA}) = \mathbf{E}\left[\,\mathbf{Adv}^{\mathrm{ds}}(\mathcal{SE}_S; \mathsf{ChA})\right].$$

Similarly, one can extend the definition of semantic security to the setting of seeded encryption.

Note that in the special case of one individual seed value, we obtain the special case of unseeded encryption given above.

FROM SEEDED TO UNSEEDED ENCRYPTION. We discuss how to generically transform any seeded encryption function into a conventional (seedless) encryption function. This transformation is rate preserving, i.e., the rate of the resulting scheme is (asymptotically) the same as the one of the original seeded encryption scheme.

The main idea behind our construction – which we call **SR** (for Seed Recycle) – is to encrypt multiple message blocks using the underlying seeded encryption scheme with the *same* seed, and combine the resulting encryptions in one single ciphertext. An error-corrected version of the seed is included the ciphertext to ensure decryption. If sufficiently many encryptions are combined into one ciphertext, the

**Transform** $\mathcal{E}(M)$: // $M \in \{0,1\}^m$

$S \leftarrow\$ \; \textsc{Sds}$
$M[1], \ldots, M[t] \xleftarrow{b} M$
For $i = 1$ to $t$ do
  $C[i] \leftarrow\$ \; \mathcal{SE}(S, M[i])$
Ret $\mathsf{E}(S) \,\|\, C[1] \,\|\, \cdots \,\|\, C[t]$ .

**Transform** $\mathcal{D}(C_0 \| C_1)$: // $C_0 \in \{0,1\}^{\ell_0}, C_1 \in \{0,1\}^{t\ell_1}$

$C_1[1], \ldots, C_1[t] \xleftarrow{\ell_1} C_1$
$S \leftarrow \mathsf{D}(C_0)$
For $i = 1$ to $t$ do
  $M[i] \leftarrow \mathcal{SD}(S, C[i])$
Ret $M[1] \,\|\, \cdots \,\|\, M[t]$ .

Figure 2: **Encryption from a seeded encryption.** Encryption function $\mathcal{E} = \mathbf{SR}[\mathcal{SE}, \mathsf{E}]$, and associated decryption function $\mathcal{D}$ for the channel $\mathsf{ChR} = \mathsf{ChR}_0 \,\|\, \mathsf{ChR}_1^t$, where $\mathsf{ChR}_0 : \{0,1\}^e \to \{0,1\}^{\ell_0}$ and $\mathsf{ChR}_1 : \{0,1\}^n \to \{0,1\}^{\ell_1}$. By $X[1], \ldots, X[c] \xleftarrow{b} X$ we mean that $bc$-bit string $X$ is split into $b$-bit blocks.

cost of including the seed is asymptotically vanishing, hence preserving the rate of the underlying seeded encryption. Moreover, no privacy must be guaranteed for the seed, as it can be made public, provided the underlying seeded encryption is DS-secure.

More concretely, let $\mathcal{SE} : \textsc{Sds} \times \{0,1\}^b \to \{0,1\}^n$ be a seeded encryption function and let $\mathsf{E} : \textsc{Sds} \to \{0,1\}^e$ be an efficiently computable injective function. For a parameter $t \geq 1$, the encryption function $\mathcal{E} = \mathbf{SR}_t[\mathcal{SE}, \mathsf{E}]$ takes as input a message $M \in \{0,1\}^m$, where $m = t \cdot b$, and splits it into $t$ $b$-bit blocks $M[1], \ldots, M[t]$. It selects a random seed $S \leftarrow\$ \; \textsc{Sds}$, and then encrypts the individual message blocks as $C[i] = \mathcal{SE}(S, M[i])$. The final $(e + t \cdot n)$-bit ciphertext consists of the concatenation of $C[0] = \mathsf{E}(S)$ and $C[1], \ldots, C[t]$. The encryption function $\mathcal{E}$ is described in Figure 2 for completeness.

Decryption for **SR**. First, recall that a *code* is an injective function $\mathsf{E} : \{0,1\}^k \to \{0,1\}^e$ for $k \leq e$. Given a channel $\mathsf{ChR} : \{0,1\}^e \to \{0,1\}^\ell$, a decoder for $\mathsf{E}$ over $\mathsf{ChR}$ is a an algorithm $\mathsf{D} : \{0,1\}^\ell \to \{0,1\}^k$. As in the case of decryption, its decoding error is defined as $\mathbf{DE}(\mathsf{E}; \mathsf{D}; \mathsf{ChR}) = \max_{M \in \{0,1\}^k} \Pr[\, \mathsf{D}(\mathsf{ChR}(\mathsf{E}(M))) \neq M \,]$.

We assume that $\mathcal{E}$ is used over a channel $\mathsf{ChR} = \mathsf{ChR}_0 \,\|\, \mathsf{ChR}_1^t$ which operates by independently processing the first $n$ ciphertext bits through a channel $\mathsf{ChR}_0 : \{0,1\}^e \to \{0,1\}^{\ell_0}$ and each subsequent $n$-bit block is sent (independently) through a channel $\mathsf{ChR}_1 : \{0,1\}^n \to \{0,1\}^{\ell_1}$. The goal of the function $\mathsf{E}$ is to operate as a code ensuring recovery of the seed. Therefore, for any function $\mathsf{D} : \{0,1\}^{\ell_0} \to \textsc{Sds}$, and decryption function $\mathcal{SD} : \textsc{Sds} \times \{0,1\}^{\ell_1} \to \{0,1\}^n$ for $\mathcal{SE}$ over $\mathsf{ChR}_1$, we specify the corresponding decryption function $\mathcal{D}$ for $\mathcal{E}$ over $\mathsf{ChR}$ as in Figure 2. The following lemma summarizes the relation between its decryption error and the ones of $\mathsf{D}$ and $\mathcal{SE}$, and its proof follows by a simple union bound.

**Lemma 4.1 [Correct decryption of SR]** *Let $t \geq 1$, and let $\mathsf{ChR} = \mathsf{ChR}_0 \,\|\, \mathsf{ChR}_1^t$ be such that $\mathsf{ChR}_0 : \{0,1\}^e \to \{0,1\}^{\ell_0}$ and $\mathsf{ChR}_1 : \{0,1\}^n \to \{0,1\}^{\ell_1}$. Moreover, let $\mathcal{SE} : \textsc{Sds} \times \{0,1\}^b \to \{0,1\}^n$, $\mathcal{SD} : \textsc{Sds} \times \{0,1\}^{\ell_1} \to \{0,1\}^b$, $\mathsf{E} : \textsc{Sds} \to \{0,1\}^e$, and $\mathsf{D} : \{0,1\}^{\ell_0} \to \textsc{Sds}$. Then, for $\mathcal{E} = \mathbf{SR}_t[\mathcal{SE}, \mathsf{E}]$ and the associated decryption function $\mathcal{D}$ as above using $\mathsf{D}$,*

$$\mathbf{DE}(\mathcal{E}; \mathcal{D}; \mathsf{ChR}) \leq \mathbf{DE}(\mathsf{E}; \mathsf{D}; \mathsf{ChR}_0) + t \cdot \mathbf{DE}(\mathcal{SE}; \mathcal{SD}; \mathsf{ChR}_1) \; . \; \blacksquare$$

We note that Lemma 4.1 can be extended to the case where the channels $\mathsf{ChR}_0$, as well as the $t$ usage of $\mathsf{ChR}_1$, are not necessarily independent, provided they do behave individually as $\mathsf{ChR}_0$ and $\mathsf{ChR}_1$, respectively.

Security of **SR**. We now turn to proving that DS security of $\mathcal{E} = \mathbf{SR}_t[\mathcal{SE}, \mathsf{E}]$ can be reduced to DS-security of $\mathcal{SE}$, at the cost of only a factor $t$ loss in the security reduction.

**Lemma 4.2 [Security of SR]** *Let $t \geq 1$, and let $\mathsf{ChA} = \mathsf{ChA}_0 \,\|\, \mathsf{ChA}_1^t$ be such that $\mathsf{ChA}_0 : \{0,1\}^e \to \{0,1\}^{\ell'_0}$ and $\mathsf{ChA}_1 : \{0,1\}^n \to \{0,1\}^{\ell'_1}$. Moreover, let $\mathcal{SE} : \textsc{Sds} \times \{0,1\}^b \to \{0,1\}^n$, $\mathsf{E} : \textsc{Sds} \to \{0,1\}^e$, and $\mathcal{E} = \mathbf{SR}_t[\mathcal{SE}, \mathsf{E}]$. Then,*

$$\mathbf{Adv}^{\mathrm{ds}}(\mathcal{E}; \mathsf{ChA}_0 \,\|\, \mathsf{ChA}_1^t) \leq t \cdot \mathbf{Adv}^{\mathrm{ds}}(\mathcal{SE}; \mathsf{ChA}_1) \; . \; \blacksquare$$

**Transform** $\mathsf{F}_i(X, S)$: $\quad$ // $X \in \{0,1\}^{\ell'_1}$, $S \in \mathrm{SDS}$

$Z[0] \leftarrow\!\!\$\ \mathsf{ChA}_0(\mathsf{E}(S))$

$\overline{M}_i[1], \ldots, \overline{M}_i[t] \stackrel{b}{\leftarrow} \overline{M}_i$

$Z[i] \leftarrow X$

For $j = 1$ to $t$, $j \neq i$, do

$\quad C[j] \leftarrow\!\!\$\ \mathcal{SE}(S, \overline{M}_i[j])$; $Z[j] \leftarrow\!\!\$\ \mathsf{ChA}_1(C[j])$

Ret $Z[0] \,\|\, Z[1] \,\|\, \cdots \,\|\, Z[t]$ .

Figure 3: **Proof of Lemma 4.2**. Description of the transform $\mathsf{F}_i$.

**Proof:** The proof proceeds by a hybrid argument. To start with, let us fix two arbitrary $m$-bit messages $M_0, M_1$. Recall that $m = t \cdot b$. For all $i \in [0 \ldots t]$, let $\overline{M}_i \in \{0,1\}^m$ be such that for all $j \in [1 \ldots t]$, the $j$-th $b$-bit block $\overline{M}_i[j]$ equals $M_1[j]$ if $j \leq i$, and $M_0[j]$ otherwise. In particular, $\overline{M}_0 = M_0$ and $\overline{M}_t = M_1$. We let $\mathsf{X}_i = \mathsf{ChA}(\mathcal{E}(\overline{M}_i))$, and by the triangle inequality,

$$\mathbf{SD}(\mathsf{ChA}(\mathcal{E}(M_0)); \mathsf{ChA}(\mathcal{E}(M_1))) = \mathbf{SD}(\mathsf{X}_0; \mathsf{X}_t) \leq \sum_{i=1}^{t} \mathbf{SD}(\mathsf{X}_{i-1}; \mathsf{X}_i) \ .$$

For convenience, let us introduce the shorthand $\mathsf{Z}(S, M) = \mathsf{ChA}_1(\mathcal{SE}(S, M))$. For all $i \in [1 \ldots t]$, we introduce the transform $\mathsf{F}_i : \{0,1\}^{\ell'_1} \times \mathrm{SDS} \to \{0,1\}^{\ell'_0 + t \cdot \ell'_1}$ described in Figure 3. Then, it is easy to verify that the outputs of $\mathsf{F}_i(\mathsf{Z}(\mathsf{S}, M_0[i]), \mathsf{S})$ and $\mathsf{F}_i(\mathsf{Z}(\mathsf{S}, M_1[i]), \mathsf{S})$ are distributed as $\mathsf{X}_{i-1}$ and $\mathsf{X}_i$, respectively. Hence:

$$\begin{aligned}
\mathbf{SD}(\mathsf{X}_{i-1}; \mathsf{X}_i) &= \mathbf{SD}(\mathsf{F}_i(\mathsf{Z}(\mathsf{S}, M_0[i]), \mathsf{S}); \mathsf{F}_i(\mathsf{Z}(\mathsf{S}, M_1[i]), \mathsf{S})) \\
&\leq \mathbf{SD}((\mathsf{Z}(\mathsf{S}, M_0[i]), \mathsf{S}); (\mathsf{Z}(\mathsf{S}, M_1[i]), \mathsf{S})) \\
&\leq \max_{M'_0, M'_1 \in \{0,1\}^b} \mathbf{SD}((\mathsf{Z}(\mathsf{S}, M'_0), \mathsf{S}); (\mathsf{Z}(\mathsf{S}, M'_1), \mathsf{S})) \\
&= \max_{M'_0, M'_1 \in \{0,1\}^b} \mathbf{E}_{S \leftarrow\$\, \mathrm{SDS}} \left[ \mathbf{SD}(\mathsf{Z}(S, M'_0); \mathsf{Z}(S, M'_1)) \right] \\
&\leq \mathbf{E}_{S \leftarrow\$\, \mathrm{SDS}} \left[ \max_{M'_0, M'_1 \in \{0,1\}^b} \mathbf{SD}(\mathsf{Z}(S, M'_0); \mathsf{Z}(S, M'_1)) \right] \\
&= \mathbf{Adv}^{\mathrm{ds}}(\mathcal{SE}; \mathsf{ChA}_1) \ ,
\end{aligned}$$

where the first inequality follows from the fact that $\mathbf{SD}(g(\mathsf{X}); g(\mathsf{Y})) \leq \mathbf{SD}(\mathsf{X}; \mathsf{Y})$ for all random variables $\mathsf{X}, \mathsf{Y}$, and all functions $g$. The final bound follows by maximizing over all $M_0, M_1 \in \{0,1\}^m$. ▮

An encryption scheme and its rate. In the asymptotic setting, we construct an encryption scheme $\overline{\mathcal{E}} = \{\mathcal{E}_k\}_{k \in \mathbb{N}}$ using the **SR** construction as follows: We start from an arbitrary seeded encryption scheme $\overline{\mathcal{SE}} = \{\mathcal{SE}_k\}_{k \in \mathbb{N}}$ such that $\mathcal{SE}_k : \mathrm{SDS}_k \times \{0,1\}^{b(k)} \to \{0,1\}^{n(k)}$, as well as from family of injective functions $\overline{\mathsf{E}} = \{\mathsf{E}_k\}_{k \in \mathbb{N}}$ with $\mathsf{E}_k : \mathrm{SDS}_k \to \{0,1\}^{e(k)}$. Also let $t : \mathbb{N} \to \mathbb{N}$ be a function such that $e(k) = o(n(k) \cdot t(k))$. Often, letting $t(k) = O(\log(n))$ will be sufficient. Then, for all $k \in \mathbb{N}$, the encryption algorithm $\mathcal{E}_k : \{0,1\}^{t(k) \cdot b(k)} \to \{0,1\}^{e(k) + t(k) \cdot n(k)}$ is defined as $\mathcal{E}_k = \mathbf{SR}_t[\mathcal{SE}_k, \mathsf{E}_k]$.

We conclude by verifying that the rates of $\overline{\mathcal{E}}$ and $\overline{\mathcal{SE}}$ are indeed equal:

$$\begin{aligned}
\mathbf{Rate}(\overline{\mathcal{E}}) &= \lim_{k \to \infty} \frac{t(k) \cdot b(k)}{e(k) + t(k) \cdot n(k)} \\
&= \lim_{k \to \infty} \frac{b(k)}{e(k)/t(k) + n(k)} \\
&= \lim_{k \to \infty} \frac{b(k)}{n(k)} \cdot \lim_{k \to \infty} \frac{1}{1 + e(k)/(t(k) \cdot n(k))} \\
&= \mathbf{Rate}(\overline{\mathcal{SE}}) \cdot \frac{1}{\underbrace{1 + \lim_{k \to \infty} e(k)/(t(k) \cdot n(k))}_{=0}} = \mathbf{Rate}(\overline{\mathcal{SE}}) \ .
\end{aligned}$$

# 5 A DS-Secure Scheme Achieving Secrecy Capacity

In this section, we turn to our main technical result, a seeded encryption scheme achieving DS-security. Its rate, for a large set of adversary channels, is optimal, meaning it equals the secrecy capacity. Using the **SR** construction from the previous section, our scheme yields an unseeded encryption scheme with optimal rate.

## 5.1 The ItE Construction

In the following, we present our generic construction of an encryption function, which we call **ItE** (Invert-then-Encode). Before giving any details, however, let us start with the high-level idea underlying our approach. For simplicity, let us focus on the case where $\mathsf{ChR}$ and $\mathsf{ChA}$ are BSC's with respective crossover probabilities $p_R < p_A \leq 1/2$. Let us also assume the goal is the simpler one of Alice and Bob agreeing on an $n$-bit key rather than transmitting a message. If we let the seed $S \in \text{SDS}$ be the seed for an extractor $\mathsf{Ext}\colon \text{SDS} \times \{0,1\}^k \to \{0,1\}^m$ and given an error-correcting code $\mathsf{E}\colon \{0,1\}^k \to \{0,1\}^n$ for reliable communication over $\mathsf{BSC}_{p_R}$, a natural approach consists of Alice sending $\mathsf{E}(R)$, for a random $k$-bit $R$, to Bob, via $\mathsf{ChR}$, and both parties now derive the key as $K = \mathsf{Ext}(S, R)$.

Proving that this approach works requires estimating $\mathbf{H}_\infty(R|Z) = -\lg(\sum_z \max_r \Pr[R = r, Z = z])$, where $Z = \mathsf{BSC}^n_{p_A}(\mathsf{E}(R))$ is the information received by Eve. Yet, it is not hard to see that that the most likely outcome, when $Z = z$, is that $R$ equals the unique $r$ such that $\mathsf{E}(r) = z$, and that hence $\mathbf{H}_\infty(R|Z) = n \cdot \lg(1/(1-p_A))$, which is smaller than $h_2(p_A) - h_2(p_R)$, and which also upper bounds the length of the derived key $K$. To overcome this, we will observe the following: We can think of $\mathsf{BSC}_{p_A}$ as adding an $n$-bit vector $E$ to its input $\mathsf{E}(R)$, where each bit $E[i]$ of the noise vector takes value one with probability $p_A$. With overwhelming probability, $E$ is (roughly) uniformly distributed on the set of $n$-bit vectors with hamming weight (approximately) $p_A \cdot n$ and there are (approximately) $2^{n \cdot h_2(p_A)}$ such vectors. Therefore, choosing the noise uniformly from such vectors does not change the experiment much, and moreover, in this new experiment, one can show that roughly $\mathbf{H}_\infty(R|Z) \geq k - n \cdot (1 - h_2(p_A))$. We will make this precise for a general class of symmetric channels via the notion of *smooth min-entropy* [29].

While both $\mathsf{Ext}$ and $\mathsf{E}$ can be instantiated so that the secret-key rate satisfies $|K|/n \approx h_2(p_A) - h_2(p_R)$, which is the secrecy capacity, recall that our goal is way more ambitious: Alice wants to send an *arbitrary message of her choice*. The obvious way to do this is obtain a key $K$ as above and then send $K \oplus M$. But this at least halves the rate, which becomes far from optimal. Our approach instead is to use an extractor $\mathsf{Ext}$ that is invertible in the sense that given $M$ and $S$, we can sample a random $R$ such that $\mathsf{Ext}(S, R) = M$. We then encrypt a message $M$ as $\mathsf{E}(R)$, where $R$ is a random preimage of $M$ under $\mathsf{Ext}(S, \cdot)$. However, the above argument only yields, at best, security for a randomly chosen input. In contrast, showing DS-security accounts to proving, for any two messages $M_0$ and $M_1$, that $\mathsf{BSC}^n_{p_A}(\mathsf{E}(R_0))$ and $\mathsf{BSC}^n_{p_A}(\mathsf{E}(R_1))$ are statistically close, where $R_i$ is uniform such that $\mathsf{Ext}(S, R_i) = M_i$. To make things even worse, the messages $M_0$ and $M_1$ are allowed to depend on the seed. The main challenge is that such proof appears to require detailed knowledge of the combinatorial structure of $\mathsf{E}$ and $\mathsf{Ext}$. In particular, we remark that it is not possible to provide a direct proof that the encryption of an arbitrary message is uniformly distributed, even in the simpler case where the message is seed-independent. In fact, for most codes, ciphertexts turn out not to be uniform at all.

Instead, we will take a completely different approach: We prove a general result, of independent interest, which shows that any seeded encryption function with appropriate linearity properties is DS-secure whenever it is secure for randomly chosen inputs. This result is surprising, as random-input security does *not*, in general, imply chosen-input security. A careful choice of the extractor to satisfy these requirements, combined with the above idea, will hence yield a scheme achieving DS-security.

(Invertible) Extractors. A function $\mathsf{Ext}\colon \text{SDS} \times \{0,1\}^k \to \{0,1\}^b$ is called a $(h, \alpha)$-*extractor* (strong, average case extractor in the terminology of [12]) if $\mathbf{SD}((\mathsf{Ext}(\mathsf{S}, \mathsf{X}), \mathsf{Z}, \mathsf{S}); (\mathsf{U}, \mathsf{Z}, \mathsf{S})) \leq \alpha$ for all pairs of (correlated) random variables $(\mathsf{X}, \mathsf{Z})$ over $\{0,1\}^k \times \{0,1\}^*$ with $\mathbf{H}_\infty(\mathsf{X}|\mathsf{Z}) \geq h$, where additionally $\mathsf{S}$ and $\mathsf{U}$ are uniform on SDS and $\{0,1\}^b$, respectively. We will say that $\mathsf{Ext}$ is *regular* if for all $S \in \text{SDS}$, the function $\mathsf{Ext}(S, \cdot)$ is regular, meaning every point in the range has the same number of preimages.

| **Transform** $\mathcal{SE}(S, M)$: // $S \in \text{Sds}$, $M \in \{0,1\}^b$ | **Transform** $\mathcal{SD}(S, C)$: // $C \in \{0,1\}^\ell$ |
|---|---|
| $R \leftarrow_\$ \{0,1\}^r$ | $X \leftarrow \mathsf{D}(C)$ |
| $X \leftarrow \mathsf{Inv}(S, R, M)$ | $M \leftarrow \mathsf{Ext}(S, X)$ |
| Ret $\mathsf{E}(X)$ . | Ret $M$ . |

Figure 4: **Seeded Encryption function $\mathcal{SE} = \mathbf{ItE}[\mathsf{Inv}, \mathsf{E}]$ and associated decryption function $\mathcal{SD}$.**

Recall that a function $H\colon \text{Sds} \times \{0,1\}^k \to \{0,1\}^b$ is *two-universal* if $\Pr[H(S, X) = H(S, X')] \leq 2^{-m}$ for all distinct $X, X' \in \{0,1\}^k$ when $S \leftarrow_\$ \text{Sds}$. The following average-case version of the Leftover Hash Lemma (LHL) of [16], due to [12], implies that a two-universal function is an essentially-optimal extractor:

**Lemma 5.1** *Let* $H\colon \text{Sds} \times \{0,1\}^k \to \{0,1\}^b$ *be a two-universal function. Let* $\mathsf{S}$ *be uniform over* Sds. *Let* $\mathsf{X}, \mathsf{Z}$ *be random variables over* $\{0,1\}^k$ *and* $\{0,1\}^*$ *respectively, and let* $\mathsf{U}$ *be uniform on* $\{0,1\}^b$, *independent of* $\mathsf{X}$, $\mathsf{Z}$ *and* $\mathsf{S}$. *Then,*

$$\mathbf{SD}((H(\mathsf{S}, \mathsf{X}), \mathsf{Z}, \mathsf{S}); (\mathsf{U}, \mathsf{Z}, \mathsf{S})) \leq \frac{1}{2}\sqrt{2^{b - \mathbf{H}_\infty(\mathsf{X}|\mathsf{Z})}} \ . \ \blacksquare$$

Specifically, this says that $H$ is a $(h, \alpha)$-extractor with $h = b - 2 - 2\lg\alpha$.

Our approach will rely on extractors which can efficiently be inverted. We say that a function $\mathsf{Inv} : \text{Sds} \times \{0,1\}^r \times \{0,1\}^b \to \{0,1\}^k$ is an *inverter* for an extractor $\mathsf{Ext} : \text{Sds} \times \{0,1\}^k \to \{0,1\}^b$ if for all $S \in \text{Sds}$ and $Y \in \{0,1\}^b$, and for $\mathsf{R}$ uniform over $\{0,1\}^k$, the random variable $\mathsf{Inv}(S, \mathsf{R}, Y)$ is uniformly distributed on $\{\, X \in \{0,1\}^k \ : \ \mathsf{Ext}(S, X) = Y \,\}$, the set of preimages of $Y$ under $\mathsf{Ext}(S, \cdot)$.

To make this concrete we give an example of an extractor with an efficiently computable inverter. Recall that $k$-bit strings can be interpreted as elements of the finite field $\mathrm{GF}(2^k)$, allowing us to define a multiplication operator $\odot$ on $k$-bit strings. Then, for $\text{Sds} = \{0,1\}^k \setminus 0^k$, we consider the function $\mathsf{Ext} : \{0,1\}^k \times \{0,1\}^k \to \{0,1\}^b$ which, on inputs $S \in \text{Sds}$ and $X \in \{0,1\}^k$, outputs the first $b$ bits of $X \odot S$. It is easy to see that $\mathsf{Ext}$ is regular if $0^k$ is removed from the set of seeds. In Appendix B we prove the following using the LHL:

**Lemma 5.2** *For all* $\alpha \in (0, 1]$ *and all* $b \leq k - 2\lg(1/\alpha) - 2$, *the function* $\mathsf{Ext}$ *is a* $(b + 2\lg(1/\alpha) + 2, \alpha)$-*extractor.*

An efficient inverter $\mathsf{Inv} : \text{Sds} \times \{0,1\}^{k-b} \times \{0,1\}^b \to \{0,1\}^k$ is obtained by letting $\mathsf{Inv}(S, R, M) = S^{-1} \odot (M \,\|\, R)$ where $S^{-1}$ is the inverse of $S$ with respect to multiplication in $\mathrm{GF}(2^k)$. Invertible extractors were used in [7] but their setting was much simpler than ours and they achieve only security for random inputs.

Encryption. We now describe the seeded encryption function of **ItE**. In the following, let $\mathsf{Ext} : \text{Sds} \times \{0,1\}^k \to \{0,1\}^b$ be a regular extractor with inverter $\mathsf{Inv} : \text{Sds} \times \{0,1\}^r \times \{0,1\}^b \to \{0,1\}^k$. Also let $\mathsf{E} : \{0,1\}^k \to \{0,1\}^n$ be a function with $k \leq n$, later to be instantiated via an appropriate error-correcting code. The encryption function $\mathcal{SE} = \mathbf{ItE}[\mathsf{Inv}, \mathsf{E}]$ is described in Figure 4: It applies the extractor inverter (with fresh randomness $R$) to the message $M$ and the seed $S$ to obtain an intermediate value $X = \mathsf{Inv}(S, R, M)$, which is then encoded using $\mathsf{E}$ to obtain the ciphertext.

Decryption. Given a channel $\mathsf{ChR} : \{0,1\}^n \to \{0,1\}^\ell$, the goal of the function $\mathsf{E}$ in $\mathcal{SE}$ above is to operate as an error-correcting code ensuring decryptability of the generated ciphertexts. Therefore, for any function $\mathsf{D} : \{0,1\}^\ell \to \{0,1\}^b$, we can define the corresponding decryption function $\mathcal{SD}$ for $\mathcal{SE}$ over $\mathsf{ChR}$ as in Figure 4. The following lemma summarizes the relation between its decryption error and the one of $\mathsf{D}$.

**Lemma 5.3 [Correct decryption]** *Let* $\mathsf{ChR} : \{0,1\}^n \to \{0,1\}^\ell$ *be a channel, and let* $\mathcal{SE}$, $\mathcal{SD}$, $\mathsf{E}$, *and* $\mathsf{D}$ *be as above. Then,* $\mathbf{DE}(\mathcal{SE}; \mathcal{SD}; \mathsf{ChR}) \leq \mathbf{DE}(\mathsf{E}; \mathsf{D}; \mathsf{ChR})$. $\blacksquare$

Security. Below, we discuss the security of **ItE**. Our approach consists of two steps: We first introduce a metric based on statistical distance capturing the random-message security of a seeded encryption function, and prove random-message security of **ItE**. Subsequently, we prove a general result showing that random-message security implies DS security in many scenarios, and apply it to **ItE**.

## 5.2 Random-Message Security of ItE

Random distinguishing-security. We now address the problem of proving security of **ItE** under *random* messages. To this end, we introduce a new security metric rds based on the statistical distance. Specifically, for a seeded encryption function $\mathcal{SE} : \text{Sds} \times \{0,1\}^b \to \{0,1\}^n$ and a channel $\mathsf{ChA} : \{0,1\}^n \to \{0,1\}^*$, we define the rds advantage as

$$\mathbf{Adv}^{\mathrm{rds}}(\mathcal{SE}; \mathsf{ChA}) = \mathbf{E}\left[\mathbf{SD}((\mathsf{ChA}(\mathcal{SE}(S, \mathsf{U})), \mathsf{U}); (\mathsf{ChA}(\mathcal{SE}(S, \mathsf{U}')), \mathsf{U}))\right] ,$$

where $\mathsf{U}$ and $\mathsf{U}'$ are independent $b$-bit inputs, and the expectation is taken over the choice of the seed $S$. Below, we will prove that, surprisingly, RDS-security is often sufficient in order to infer DS-security of a seeded encryption function.

RDS-Security for **ItE**. We now want to prove an upper bound on rds advantage for $\mathcal{SE} = \mathbf{ItE}[\mathsf{Inv}, \mathsf{E}]$, where $\mathsf{Inv}$ is the inverter of a regular extractor $\mathsf{Ext}$.

It is crucial to remark that the joint distribution of $(\mathsf{X}, \mathsf{U})$ is identical if we (i) sample a uniform random $b$-bit message $\mathsf{U}$, a random $r$-bit string $\mathsf{R}$, and compute $\mathsf{X} \leftarrow \mathsf{Inv}(S, \mathsf{R}, \mathsf{U})$, or if instead (ii) we pick $\mathsf{X}$ uniformly at random, and then compute $\mathsf{U} \leftarrow \mathsf{Ext}(\mathsf{S}, \mathsf{X})$. But then, intuitively, we expect that $\mathsf{X}$, when $\mathsf{E}(\mathsf{X})$ is sent through $\mathsf{ChA}$, has sufficiently high min-entropy $h$ in the eyes of the adversary, hence implying that if $\mathsf{Ext}$ is an $(h, \alpha)$-extractor,

$$\mathbf{Adv}^{\mathrm{rds}}(\mathcal{SE}; \mathsf{ChA}) = \mathbf{E}\left[\mathbf{SD}((\mathsf{ChA}(\mathcal{SE}(S, \mathsf{U})), \mathsf{U}); (\mathsf{ChA}(\mathcal{SE}(S, \mathsf{U}')), \mathsf{U}))\right] \leq \alpha ,$$

since for any two transforms $T_1, T_2$, $\mathbf{SD}((T_1(\mathsf{S}), \mathsf{S}); (T_2(\mathsf{S}), \mathsf{S})) = \mathbf{E}\left[\mathbf{SD}(T_1(S); T_2(S))\right]$, where the expectation is over the choice of $S$ according to $P_{\mathsf{S}}$. In order to lower bound the entropy $\mathbf{H}_\infty(\mathsf{X}|\mathsf{ChA}(\mathsf{E}(\mathsf{X})))$ we will use the following observation: For a symmetric channel $\mathsf{ChA}$, let $\mathbf{H}(\mathsf{ChA}) = \mathbf{H}(\mathsf{ChA}(X))$ for *any* input $X$. (The entropy $\mathbf{H}(\mathsf{ChA}(X))$ is the same, regardless of the input, since the *rows* of the transition probability matrix of $\mathsf{ChA}$ are all permutations of each other.) Then, we are going to prove that if we use the channel $n$ times, each time to transmit a bit, then we can always see $\mathsf{ChA}$ as adding some noise whose distribution is *statistically close* to a noise distribution where all values are taken with probability at most $2^{-n\mathbf{H}(\mathsf{ChA})}$. This is formalized via the notion of *$\epsilon$-smooth min-entropy* [29] of a distribution $P$, defined as

$$\mathbf{H}_\infty^\epsilon(P) = \max_{Q : \mathbf{SD}(P;Q) \leq \epsilon} \mathbf{H}_\infty(Q) .$$

Analogously, we define $\mathbf{H}_\infty^\epsilon(\mathsf{X}) = \mathbf{H}_\infty^\epsilon(P_{\mathsf{X}})$ for every random variable $\mathsf{X}$ with distribution $P_{\mathsf{X}}$. The following lemma, first shown by Holenstein and Renner [18] in a more general setting, states that the smooth min-entropy of multiple independent samples is, on average, nearly as large as the Shannon entropy of an individual sample.

**Lemma 5.4 [18]** *Let $\mathsf{X}_1, \ldots, \mathsf{X}_n$ be independent samples from a distribution $P$ on a finite set $X$, and let $\delta > 0$. Then, $\mathbf{H}_\infty^\epsilon(\mathsf{X}_1 \ldots \mathsf{X}_n) \geq n \cdot \mathbf{H}(P) - n \cdot \delta$, where*

$$\epsilon = \epsilon(\delta, n, |X|) = 2^{-\frac{n\delta^2}{2\lg^2(|X|+3)}} . \blacksquare$$

The RDS-security of $\mathcal{SE}$ is then summarized by the following lemma. Interestingly, the lemma does not need any assumption on $\mathsf{E}$, other than the fact that it is injective.

**Lemma 5.5 [RDS-security of ItE]** *Let $\delta > 0$ and $\text{OutA} \subseteq \{0,1\}^*$. Also, let $\mathsf{ChA} : \{0,1\} \to \text{OutA}$ be a symmetric channel and let $\mathcal{SE} = \mathbf{ItE}[\mathsf{Inv}, \mathsf{E}]$, where $\mathsf{Inv}$ is the inverter of a regular $(k - n \cdot (\lg(|\text{OutA}|) - \mathbf{H}(\mathsf{ChA}) + \delta), \alpha)$-extractor, and $\mathsf{E}$ is injective. Then,*

$$\mathbf{Adv}^{\mathrm{rds}}(\mathcal{SE}; \mathsf{ChA}^n) \leq 2 \cdot 2^{-\frac{n\delta^2}{2\lg^2(|\text{OutA}|+3)}} + \alpha . \blacksquare$$

| Process $\Pi_1$: | Process $\Pi_2$: | Process $\Pi_3$: |
|---|---|---|
| $M \leftarrow\!\$ \{0,1\}^b$; <br> $S \leftarrow\!\$ \text{SDS}$; $R \leftarrow\!\$ \{0,1\}^r$ <br> $X \leftarrow \mathsf{Inv}(S,R,M)$; <br> $C \leftarrow \mathsf{E}(X)$ <br> $Z \leftarrow\!\$ \mathsf{ChA}^n(C)$ <br> Ret $(Z,M,S)$. | $M \leftarrow\!\$ \{0,1\}^b$; <br> $S \leftarrow\!\$ \text{SDS}$; $R \leftarrow\!\$ \{0,1\}^r$ <br> $X \leftarrow \mathsf{Inv}(S,R,M)$; <br> $C \leftarrow \mathsf{E}(X)$ <br> $Y \leftarrow\!\$ \mathsf{ChA}^n(0^n)$; $Z \leftarrow \pi_C(Y)$ <br> Ret $(Z,M,S)$. | $S \leftarrow\!\$ \text{SDS}$; $X \leftarrow\!\$ \{0,1\}^k$ <br> $M \leftarrow \mathsf{Ext}(S,X)$; <br> $C \leftarrow \mathsf{E}(X)$ <br> $Y \leftarrow\!\$ \mathsf{ChA}^n(0^n)$; $Z \leftarrow \pi_C(Y)$ <br> Ret $(Z,M,S)$. |
| **Process $\Pi_4$:** | **Process $\Pi_5$:** | **Process $\Pi_6$:** |
| $S \leftarrow\!\$ \text{SDS}$; $X \leftarrow\!\$ \{0,1\}^k$ <br> $M \leftarrow \mathsf{Ext}(S,X)$; <br> $C \leftarrow \mathsf{E}(X)$ <br> $Y \leftarrow\!\$ P'$; $Z \leftarrow \pi_C(Y)$ <br> Ret $(Z,M,S)$. | $S \leftarrow\!\$ \text{SDS}$; $X \leftarrow\!\$ \{0,1\}^k$ <br> $M \leftarrow \mathsf{Ext}(S,X)$; <br> $C \leftarrow \mathsf{E}(X)$ <br> $M' \leftarrow\!\$ \{0,1\}^b$ <br> $Y \leftarrow\!\$ P'$; $Z \leftarrow \pi_C(Y)$ <br> Ret $(Z,M,S)$. | $M \leftarrow\!\$ \{0,1\}^b$; $M' \leftarrow\!\$ \{0,1\}^b$ <br> $S \leftarrow\!\$ \text{SDS}$; $R \leftarrow\!\$ \{0,1\}^r$ <br> $X \leftarrow \mathsf{Inv}(S,R,M')$; <br> $C \leftarrow \mathsf{E}(X)$ <br> $Z \leftarrow\!\$ \mathsf{ChA}^n(C)$ <br> Ret $(Z,M',S)$. |

Figure 5: **Proof of Lemma 5.5.** Pseudocode description of the sequence of processes $\Pi_1$ to $\Pi_6$.

**Proof:** Recall that for two independent and uniform $b$-bit strings $\mathsf{U}$ and $\mathsf{U}'$,

$$\begin{aligned}\mathbf{Adv}^{\rm rds}(\mathcal{SE};\mathsf{ChA}^n) &= \mathbf{E}_{S \leftarrow\$ \text{SDS}}\left[\mathbf{SD}((\mathsf{ChA}^n(\mathcal{SE}(S,\mathsf{U})),\mathsf{U});(\mathsf{ChA}^n(\mathcal{SE}(S,\mathsf{U}')),\mathsf{U}))\right] \\ &= \mathbf{SD}((\mathsf{ChA}^n(\mathcal{SE}(\mathsf{S},\mathsf{U})),\mathsf{U},\mathsf{S});(\mathsf{ChA}^n(\mathcal{SE}(\mathsf{S},\mathsf{U}')),\mathsf{U},\mathsf{S})\ ,\end{aligned}$$

where $\mathsf{S}$ is chosen uniformly at random on SDS. The proof proceeds by giving a sequence of intermediate random processes, described in pseudo-code in Figure 5, which will be used to transition from the random variable $(\mathsf{ChA}^n(\mathcal{SE}(\mathsf{S},\mathsf{U})),\mathsf{U},\mathsf{S})$ to the random variable $(\mathsf{ChA}^n(\mathcal{SE}(\mathsf{S},\mathsf{U}')),\mathsf{U},\mathsf{S})$. For any two processes $\Pi_i$ and $\Pi_j$, we let $\mathbf{SD}(\Pi_i;\Pi_j)$ denote the statistical distance between their output distributions.

The first three processes $\Pi_1,\Pi_2$, and $\Pi_3$ are described on top of Figure 5. Process $\Pi_1$ samples a triple $(Z,M,S)$ according to the probability distribution of $(\mathsf{ChA}^n(\mathcal{SE}(\mathsf{S},\mathsf{U})),\mathsf{U},\mathsf{S})$. The second process $\Pi_2$ modifies the way in which $Z$ is sampled: As the channel $\mathsf{ChA}$ is symmetric, there must exist a permutation $\pi_1 : \text{OUTA} \to \text{OUTA}$ such that $\Pr[\mathsf{ChA}(B) = \pi_B(Y)] = \Pr[\mathsf{ChA}(0) = Y]$, where $\pi_0$ is the identity. Consequently, for all $X \in \{0,1\}^n$, we define $\pi_X(Y) = (\pi_{X[1]}(Y[1]),\ldots,\pi_{X[n]}(Y[n]))$, and

$$\Pr[\mathsf{ChA}^n(X) = \pi_X(Y)] = \Pr[\mathsf{ChA}^n(0^n) = Y]$$

for all $Y \in \text{OUTA}^n$. Accordingly, to implement the channel $\mathsf{ChA}^n$ on input $X$ in $\Pi_2$, we first sample $Y \leftarrow\!\$ \mathsf{ChA}^n(0^n)$, and then output $Z = \pi_X(Y)$. For the third process $\Pi_3$, assume that we invert the role of $X$ and $M$: that is, we *first* sample $X$ uniformly at random in $\{0,1\}^k$, and *then* set $M$ to equal $\mathsf{Ext}(S,X)$. By the regularity of $\mathsf{Ext}$, the output distributions of $\Pi_2$ and $\Pi_3$ are identical. Therefore, $\mathbf{SD}(\Pi_1,\Pi_2) = \mathbf{SD}(\Pi_2,\Pi_3) = \mathbf{SD}(\Pi_1,\Pi_3) = 0$.

In Process $\Pi_4$, we want to simplify the probability distribution of $\mathsf{ChA}^n(0^n)$ by computing its smooth min-entropy: Invoking Lemma 5.4 with $\mathsf{ChA}^n(0^n)$ lets us conclude that (for $\delta$ as in the lemma statement)

$$\mathbf{H}^{\epsilon}_{\infty}(\mathsf{ChA}^n(0^n)) \geq n \cdot (\mathbf{H}(\mathsf{ChA}) - \delta)$$

for $\epsilon = 2^{-\frac{n\delta^2}{2\lg^2(|\text{OUTA}|+3)}}$. Recall that this means that there exists a probability distribution $P'$ on $\text{OUTA}^n$ such that $\mathbf{SD}(P_{\mathsf{ChA}^n(0^n)};P') \leq \epsilon$ and $\mathbf{H}_{\infty}(P') \geq n \cdot (\mathbf{H}(\mathsf{ChA}) - \delta)$, or, equivalently,

$$P'(Y) \leq 2^{-n\cdot(\mathbf{H}(\mathsf{ChA})-\delta)} \tag{4}$$

for all $Y \in \text{OUTA}^n$. Accordingly, we transition from Process $\Pi_3$ to Process $\Pi_4$ by sampling $Y$ with respect to the probability distribution $P'$. Clearly, $\mathbf{SD}(\Pi_3;\Pi_4) \leq \mathbf{SD}(P_{\mathsf{ChA}^n(0^n)};P') \leq \epsilon$.

Let now $\mathsf{X}$, $\mathsf{Y}$, $\mathsf{Z}$ be random variables representing the respective choices of $X, Y$ and $Z$ in $\Pi_4$. Then,

$$\begin{aligned}
\mathbf{H}_\infty(\mathsf{X}|\mathsf{Z}) &= -\lg\left(\sum_{Z\in\text{OutA}^n}\max_{X\in\{0,1\}^k}\Pr\left[\mathsf{X}=X \wedge \mathsf{Z}=Z\right]\right)\\
&= -\lg\left(\sum_{Z\in\text{OutA}^n}\max_{X\in\{0,1\}^k}\Pr\left[\mathsf{X}=X\right]\cdot\Pr\left[\mathsf{Z}=Z \mid \mathsf{X}=X\right]\right)\\
&= -\lg\left(\sum_{Z\in\text{OutA}^n}\max_{X\in\{0,1\}^k}\Pr\left[\mathsf{X}=X\right]\cdot\Pr\left[\mathsf{Y}=\pi^{-1}_{\mathsf{E}(X)}(Z)\right]\right)\\
&= -\lg\left(\sum_{Z\in\text{OutA}^n}2^{-k}\cdot\max_{X\in\{0,1\}^k}P'(\pi^{-1}_{\mathsf{E}(T)}(Z))\right)\\
&\overset{(4)}{\geq} -\lg\left(\sum_{Z\in\text{OutA}^n}2^{-k}\cdot 2^{-n(\mathbf{H}(\mathsf{ChA})-\delta)}\right)\\
&= -\lg\left(2^{\lg(|\text{OutA}|)}\cdot 2^{-k}\cdot 2^{-n(\mathbf{H}(\mathsf{ChA})-\delta)}\right) = k - n\cdot(\lg(|\text{OutA}|)-\mathbf{H}(\mathsf{ChA})+\delta)\ .
\end{aligned}$$

In Process $\Pi_5$, instead of $M$, we then output an independent random value $M'$. Using the fact that $\mathsf{Ext}$ is a $(k - n\cdot(\lg(|\text{OutA}|)-\mathbf{H}(\mathsf{ChA})+\delta), \alpha)$-extractor, we directly obtain $\mathbf{SD}(\Pi_4;\Pi_5)\leq\alpha$. We conclude by undoing the changes we had from Process $\Pi_1$, coming back to the original choice of $X$, $C$, and $Z$, while still outputting $M'$ instead of $M$. It is easy to see that $\mathbf{SD}(\Pi_5;\Pi_6)\leq\epsilon$.

The final bound in the lemma statement follows via the triangle inequality, by adding all distances between consecutive processes. ▌

## 5.3 DS-Security of ItE

As proven above, using invertible extractors with appropriate parameters is amenable to proving RDS-security. However, proving DS-security seems to require a better grasp of the combinatorial structure of $\mathsf{Ext}$ and $\mathsf{Inv}$, as well as of the channel $\mathsf{ChA}$. Interestingly, we now show that such requirement is quite minimal as a corollary of a more general result relating RDS- and DS-security, which we now explain.

From RDS- to DS-security We first recall the following notions from [3], adapted to the more general setting of seeded encryption: Think of a randomized seeded encryption function $\mathcal{SE}:\text{Sds}\times\{0,1\}^b\to\{0,1\}^n$ as a deterministic map $\{0,1\}^r\times\text{Sds}\times\{0,1\}^b\to\{0,1\}^n$, where the first argument takes the role of the random coins. We call $\mathcal{SE}$ *separable* if

$$\mathcal{SE}(R,S,M) = \mathcal{SE}(R,S,0^b)\oplus\mathcal{SE}(0^r,S,M)$$

for all $R\in\{0,1\}^r$, $S\in\text{Sds}$, and $M\in\{0,1\}^b$. Also, $\mathcal{SE}$ is *message linear* if $\mathcal{SE}(0^r,S,\cdot):\{0,1\}^b\to\{0,1\}^n$ is linear for all $S\in\text{Sds}$.

We now state and prove the following lemma, which related RDS and DS security for seeded encryption functions when transmitting each ciphertext bit over a symmetric channel.

**Lemma 5.6 [RDS ⇒ DS]** *Let* $\text{OutA}\subseteq\{0,1\}^*$. *For any symmetric channel* $\mathsf{ChA}:\{0,1\}\to\text{OutA}$, *if* $\mathcal{SE}:\text{Sds}\times\{0,1\}^b\to\{0,1\}^n$ *is separable and message linear, then*

$$\mathbf{Adv}^{\text{ds}}(\mathcal{SE};\mathsf{ChA}^n)\leq 2\cdot\mathbf{Adv}^{\text{rds}}(\mathcal{SE};\mathsf{ChA}^n)\ .$$ ▌

An analogous version of this lemma for the simpler case of mutual-information security and unseeded encryption is given in [3]. Here, we extend their result to the setting of unseeded encryption and of DS-security. The proof of Lemma 5.6 will make use of the following technical statement from [3].

**Lemma 5.7 [3]** *Let* $\mathcal{SE}:\text{Sds}\times\{0,1\}^b\to\{0,1\}^n$ *be separable and message linear, let* $\mathsf{ChA}:\{0,1\}\to\text{OutA}$ *be a symmetric channel, and, for all* $S\in\text{Sds}$, *let* $\mathsf{Ch}_{\mathcal{SE},S}:\{0,1\}^b\to\text{OutA}^n$ *be the channel which on input* $M\in\{0,1\}^b$ *outputs* $\mathsf{ChA}(\mathcal{SE}(S,M))$. *Then,* $\mathsf{Ch}_{\mathcal{SE},S}$ *is symmetric for all* $S\in\text{Sds}$.

The proof of Lemma 5.6 centrally relies on the following fact: It states that for a symmetric channel $\mathsf{Ch}$, the statistical distance between $\mathsf{Ch}(\mathsf{U})$ for a uniform random input $\mathsf{U}$ and $\mathsf{Ch}(M)$ is the same regardless of the choice of the input $M$.

**Lemma 5.8** *Let* $\mathsf{Ch} : \{0,1\}^b \to \textsc{Out}$ *be a symmetric channel. Let* $\mathsf{U}$ *be a uniformly distributed b-bit string. Then, there exists* $\Delta(\mathsf{Ch})$ *such that* $\Delta(\mathsf{Ch}) = \mathbf{SD}(\mathsf{Ch}(\mathsf{U}); \mathsf{Ch}(M))$ *for all* $M \in \{0,1\}^b$.

**Proof:** We partition the output set $\textsc{Out} = \bigcup_{i=1}^{r} \textsc{Out}_i$ so that the sub-matrices $W[\cdot, \textsc{Out}_i]$ are strongly symmetric. By the definition of statistical distance,

$$\begin{aligned}\mathbf{SD}(\mathsf{Ch}(\mathsf{U}); \mathsf{Ch}(M)) &= \frac{1}{2} \sum_{Y \in \textsc{Out}} |\Pr[\mathsf{Ch}(\mathsf{U}) = Y] - \Pr[\mathsf{Ch}(M) = Y]| \\ &= \frac{1}{2} \sum_{i=1}^{r} \sum_{Y \in \textsc{Out}_i} |\Pr[\mathsf{Ch}(\mathsf{U}) = Y] - \Pr[\mathsf{Ch}(M) = Y]| \,.\end{aligned}$$

Fix some $i \in [1 \dots r]$. Then, for all $Y, Y' \in \textsc{Out}_i$,

$$\Pr[\mathsf{Ch}(\mathsf{U}) = Y] = \frac{1}{2^b} \sum_{M \in \{0,1\}^b} W[M, Y] = \frac{1}{2^b} \sum_{M \in \{0,1\}^b} W[M, Y'] = \Pr[\mathsf{Ch}(\mathsf{U}) = Y'] \,,$$

using the fact that the columns $W[\cdot, Y]$ and $W[\cdot, Y']$ are a permutation of each other.

Then, in particular, with $p_i = \Pr[\mathsf{Ch}(\mathsf{U}) = Y]$ for any $Y \in \textsc{Out}_i$, we can rewrite the above as

$$\mathbf{SD}(\mathsf{Ch}(\mathsf{U}); \mathsf{Ch}(M)) = \frac{1}{2} \sum_{i=1}^{r} \sum_{Y \in \textsc{Out}_i} |p_i - W[M, Y]| \,.$$

However, since the rows $W[M, \textsc{Out}_i]$ and $W[M', \textsc{Out}_i]$ are permutations of each other for any two messages $M, M' \in \{0,1\}^b$ and for all $i \in [1 \dots r]$, it follows that $\mathbf{SD}(\mathsf{Ch}(\mathsf{U}); \mathsf{Ch}(M)) = \mathbf{SD}(\mathsf{Ch}(\mathsf{U}); \mathsf{Ch}(M'))$. ▌

With both lemmas at hand, we can now turn to the proof of Lemma 5.6.

**Proof of of Lemma 5.6:** We first observe that with $\mathsf{Ch}_{\mathcal{SE},S}$ defined as in Lemma 5.7,

$$\begin{aligned}\mathbf{Adv}^{\mathrm{rds}}(\mathcal{SE}; \mathsf{ChA}^n) &= \mathbf{E}_{S \leftarrow\$ \textsc{Sds}} \left[\mathbf{SD}((\mathsf{ChA}^n(\mathcal{SE}(S, \mathsf{U})), \mathsf{U}); (\mathsf{ChA}^n(\mathcal{SE}(S, \mathsf{U}')), \mathsf{U}))\right] \\ &= \mathbf{E}_{S \leftarrow\$ \textsc{Sds}} \left[\frac{1}{2^m} \sum_{M \in \{0,1\}^m} \mathbf{SD}(\mathsf{Ch}_{\mathcal{SE},S}(\mathsf{U}'); \mathsf{Ch}_{\mathcal{SE},S}(M))\right] \\ &= \mathbf{E}_{S \leftarrow\$ \textsc{Sds}} \left[\Delta(\mathsf{Ch}_{\mathcal{SE},S})\right] \,,\end{aligned}$$

where the last equality follows from Lemma 5.8. On the other hand, by the triangle inequality,

$$\begin{aligned}\mathbf{Adv}^{\mathrm{ds}}(\mathcal{SE}; \mathsf{ChA}^n) &= \mathbf{E}_{S \leftarrow\$ \textsc{Sds}} \left[\max_{M_0, M_1 \in \{0,1\}^b} \mathbf{SD}(\mathsf{ChA}^n(\mathcal{SE}(S, M_0)); \mathsf{ChA}^n(\mathcal{SE}(S, M_1)))\right] \\ &= \mathbf{E}_{S \leftarrow\$ \textsc{Sds}} \left[\max_{M_0, M_1 \in \{0,1\}^b} \mathbf{SD}(\mathsf{Ch}_{\mathcal{SE},S}(M_0); \mathsf{Ch}_{\mathcal{SE},S}(M_1))\right] \\ &\leq \mathbf{E}_{S \leftarrow\$ \textsc{Sds}} \left[\max_{M_0, M_1 \in \{0,1\}^b} (\mathbf{SD}(\mathsf{Ch}_{\mathcal{SE},S}(M_0); \mathsf{Ch}_{\mathcal{SE},S}(\mathsf{U})) + \mathbf{SD}(\mathsf{Ch}_{\mathcal{SE},S}(\mathsf{U}); \mathsf{Ch}_{\mathcal{SE},S}(M_1)))\right] \\ &= 2 \cdot \mathbf{E}_{S \leftarrow\$ \textsc{Sds}} \left[\Delta(\mathsf{Ch}_{\mathcal{SE},S})\right] \\ &= 2 \cdot \mathbf{Adv}^{\mathrm{rds}}(\mathcal{SE}; \mathsf{ChA}^n) \,,\end{aligned}$$

which concludes the proof. ▌

DS-Security of **ItE**. Coming back to the concrete case of **ItE**, we say that an extractor-inverter $\mathsf{Inv} : \textsc{Sds} \times \{0,1\}^r \times \{0,1\}^b \to \{0,1\}^k$ is *output linear* if $\mathsf{Inv}(S, 0^r, \cdot)$ is linear for all $S \in \textsc{Sds}$. Moreover, it is *separable* if

$$\mathsf{Inv}(S, R, Y) = \mathsf{Inv}(S, R, 0^b) \oplus \mathsf{Inv}(S, 0^r, Y) \tag{5}$$

for all $S \in \textsc{Sds}$, $R \in \{0,1\}^r$, and $Y \in \{0,1\}^b$. Note that the inverter for the above extractor based on finite-field multiplication is easily seen to be output linear and separable, by the linearity of the map $Y \mapsto S^{-1} \odot X$. Therefore, if we instantiate $\mathcal{SE} = \mathbf{ItE}[\mathsf{Inv}, \mathsf{E}]$ so that $\mathsf{Inv}$ is both output linear and separable, and we let the map $\mathsf{E}$ be linear, the encryption function $\mathcal{SE}$ is easily seen to be message linear and separable. The following theorem now follows immediately by combining Lemma 5.6, and Lemma 5.5, and concludes our security analysis of **ItE**.

**Theorem 5.9 [DS-security of ItE]** *Let $\delta > 0$ and $\mathrm{OutA} \subseteq \{0,1\}^*$. Also, let $\mathsf{ChA} : \{0,1\} \to \mathrm{OutA}$ be a symmetric channel, and assume that $\mathsf{Inv}$ is the output-linear and separable inverter of regular $(k - n \cdot (\lg(|\mathrm{OutA}|) - \mathbf{H}(\mathsf{ChA}) + \delta), \alpha)$-extractor, and that $\mathsf{E} : \{0,1\}^k \to \{0,1\}^n$ is linear and injective. Then, for $\mathcal{SE} = \mathbf{ItE}[\mathsf{Inv}, \mathsf{E}]$,*

$$\mathbf{Adv}^{\mathrm{ds}}(\mathcal{SE}; \mathsf{ChA}^n) \leq 2 \left( 2 \cdot 2^{-\frac{n\delta^2}{2\lg^2(|\mathrm{OutA}|+3)}} + \alpha \right) . \blacksquare$$

Below, we use this theorem to discuss how to instantiate $\mathsf{Inv}$ and $\mathsf{E}$ to achieve secrecy capacity. Moreover, we also discuss extensions of this result to a wider set of channels.

## 5.4 Instantiating ItE

We now devise a seeded encryption scheme $\overline{\mathcal{SE}} = \{\mathcal{SE}_s\}_{s \in \mathbb{N}}$ achieving secrecy capacity for the most common case where each ciphertext bit is transmitted over the receiver channel $\mathsf{BSC}_{p_R}$ and the adversary channel $\mathsf{BSC}_{p_A}$, respectively, where $0 \leq p_R < p_A \leq \frac{1}{2}$. Note that by the above, the secrecy capacity here is $h_2(p_A) - h_2(p_R)$.[2] Using the **SR** construction above, $\overline{\mathcal{SE}}$ can be turned into an unseeded encryption scheme $\overline{\mathcal{E}}$ achieving secrecy capacity. In this case, the only known scheme [25] does not achieve security, not even against random-message adversaries. (A scheme achieving security whenever $p_R = 0$ was later given in the full version of [25], but no scheme is known for the typical case where $p_R > 0$.)

THE SCHEME. First recall that the (Shannon) capacity of a channel $\mathsf{Ch} : \{0,1\}^l \to \{0,1\}^*$ is

$$\mathbf{C}(\mathsf{Ch}) = \frac{1}{l} \max_{\mathsf{X}} \mathbf{I}(\mathsf{X}; \mathsf{Ch}(\mathsf{X})) .$$

For example, $\mathbf{C}(\mathsf{BSC}_p) = 1 - h_2(p)$. We need the following result (cf. e.g. [14] for a proof), which guarantees the existence of error-correcting codes achieving rate equal the capacity of a given channel.

**Lemma 5.10 [14]** *For any constants $l, d \geq 1$, and every channel $\mathsf{Ch} : \{0,1\}^l \to \{0,1\}^d$, there exists a family $\mathsf{E} = \{\mathsf{E}_s\}_{s \in \mathbb{N}}$ of linear codes $\mathsf{E}_s : \{0,1\}^{k(s)} \to \{0,1\}^{n(s)}$ (where $n(s)$ is a multiple of $l$), with corresponding decoding algorithms $\mathsf{D}_s : \{0,1\}^* \to \{0,1\}^{k(s)}$, such that (i) $\mathbf{DE}(\mathsf{E}_s; \mathsf{D}_s; \mathsf{Ch}^{n(s)/l}) = 2^{-\Theta(k(s))}$, (ii) $\lim_{s \to \infty} k(s)/n(s) = \mathbf{C}(\mathsf{Ch})$, and (iii) $\mathsf{E}$ and $\mathsf{D}$ are polynomial-time computable.* $\blacksquare$

To obtain our scheme via the **ItE** construction, we start with a family of codes $\{\mathsf{E}_s\}_{s \in \mathbb{N}}$ for $\mathsf{BSC}_{p_R}$ guaranteed to exist by Lemma 5.10, where $\mathsf{E}_s : \{0,1\}^{k(s)} \to \{0,1\}^{n(s)}$ and $\lim_{s \to \infty} k(s)/n(s) = 1 - h_2(p_R)$, or, equivalently, there exists $\nu$ such that $\nu(s) = o(1)$ and $k(s) = (1 - h_2(p_R) - \nu(s)) \cdot n(s)$. Then, we let $\delta(s) = (2\lg^2(5))^{1/2} \cdot n(s)^{-1/4}$ and $\alpha(s) = 2^{-n(s)^{1/2}}$, and use the finite-field based extractor $\mathsf{Ext}_s : \{0,1\}^{k(s)} \times \{0,1\}^{k(s)} \to \{0,1\}^{b(s)}$ (with the corresponding inverter $\mathsf{Inv}_s : \{0,1\}^{k(s)} \times \{0,1\}^{k(s)-b(s)} \times \{0,1\}^{b(s)} \to \{0,1\}^{k(s)}$), where

$$\begin{aligned} b(s) &= k(s) - n(s) \cdot (1 - h_2(p_A) + \delta(s)) + 2\lg(\alpha) \\ &= (h_2(p_A) - h_2(p_R) - \nu(s) - \delta(s) - 2 \cdot n(s)^{-1/2}) \cdot n(s) . \end{aligned}$$

We finally set $\mathcal{SE}_s = \mathbf{ItE}[\mathsf{Inv}_s, \mathsf{E}_s]$. With these parameters,

$$\begin{aligned} \mathbf{Adv}^{\mathrm{ds}}(\mathcal{SE}_s; \mathsf{BSC}_{p_A}^{n(s)}) &\leq 6 \cdot 2^{-\sqrt{n(s)}} \\ \mathbf{DE}(\mathcal{SE}_s; \mathcal{SD}_s; \mathsf{BSC}_{p_R}^{n(s)}) &\leq 2^{-\Theta(k(s))} \end{aligned}$$

by Theorem 5.9 and Lemma 5.3, respectively. The rate of $\mathcal{SE}_s$ is

$$\mathbf{Rate}(\mathcal{SE}_s) = h_2(p_A) - h_2(p_R) - \nu(s) - \delta(s) - \frac{2}{\sqrt{n(s)}} ,$$

[2] Recall that if $\mathsf{ChA} : \{0,1\} \to \{0,1\}^*$ and $\mathsf{ChR} : \{0,1\} \to \{0,1\}^*$ are symmetric channels, their secrecy capacity equals [22] $\mathbf{H}(\mathsf{U}|\mathsf{ChA}(\mathsf{U})) - \mathbf{H}(\mathsf{U}|\mathsf{ChR}(\mathsf{U}))$, for a uniform bit $\mathsf{U}$.

which yields

$$\mathbf{Rate}(\overline{\mathcal{SE}}) = \lim_{s\to\infty} \mathbf{Rate}(\mathcal{SE}_s) = h_2(p_A) - h_2(p_R) .$$

If we now plug $\overline{\mathcal{SE}}$ into the **SR** construction, using $t(s) = \lg(n(s))$, the resulting encryption scheme is exactly the one described in the introduction, where $A = S^{-1}$.

Some remarks. We note that it is possible to instantiate the above scheme with error-correcting codes which fall short of achieving the channel capacity, provided their rate is still larger than (roughly) $1 - h_2(p_A)$ (as otherwise $b(s)$ would become 0). In fact, this is clearly a necessary condition: A code with rate lower than $1 - h_2(p_A)$ may allow error correction when using it over $\mathsf{BSC}_{p_A}$, hence allowing the adversary to reconstruct the message.

Moreover, we point out that the same analysis can be carried out for any pair of (single input-bit) symmetric channels $\mathsf{ChR}$ and $\mathsf{ChA}$, and the resulting rate is the secrecy capacity if the capacity of $\mathsf{ChA} : \{0,1\} \to \text{OutA}$ is $\lg(|\text{OutA}|) - \mathbf{H}(\mathsf{ChA})$; this is the case if and only if a uniform input to $\mathsf{ChA}$ produces a uniform output. For other channels, our technique still yields good schemes which, however, may fall short to achieve capacity.

### 5.5 Extensions

We remark that the above presentation is constrained to single input-bit base channels for simplicity only. Our results can be extended to discrete memoryless channels with $l$-bit inputs for $l > 1$. For example, Lemma 5.5 extends to arbitrary symmetric channels $\mathsf{ChA} : \{0,1\}^l \to \text{OutA}$, at the price of replacing $n$ by $n/l$ in the security bound and in the extractor's entropy requirement. In contrast, we do not know whether Lemma 5.6 applies to arbitrary symmetric channels with $l$-bit inputs, but it does, for instance, extend to any channel of the form $\mathsf{ChA}(X) = X \oplus \mathsf{E}$, where $\mathsf{E}$ is an $l$-bit string sampled according to an input-independent noise distribution, as discussed in [3].

# A Related Work

This section surveys existing constructions of encryption schemes in the literature. Recall that given a pair of families of channels $\overline{\mathsf{ChR}} = \{\mathsf{ChR}_k\}_{k\in\mathbb{N}}$ and $\overline{\mathsf{ChA}} = \{\mathsf{ChA}_k\}_{k\in\mathbb{N}}$ (where $\mathsf{ChA}_k$ and $\mathsf{ChR}_k$ have common domain for all $k \in \mathbb{N}$), their *weak secrecy capacity* $C_w$ is the supremum of the rates achievable by pairs $(\overline{\mathcal{E}}, \overline{\mathcal{D}})$ consisting of an encryption scheme $\overline{\mathcal{E}} = \{\mathcal{E}_k\}_{k\in\mathbb{N}}$ and a decryption scheme $\overline{\mathcal{D}} = \{\mathcal{D}_k\}_{k\in\mathbb{N}}$ for $\overline{\mathcal{E}}$ and $\overline{\mathsf{ChR}}$ such that, first, the decoding requirement is satisfied, i.e.,

$$\lim_{k\to\infty} \mathbf{DE}(\mathcal{E}_k; \mathcal{D}_k; \mathsf{ChR}_k) = 0 \ .$$

and moreover,

$$\lim_{k\to\infty} \frac{\mathbf{Adv}^{\text{mis-r}}(\mathcal{E}_k; \mathsf{ChA}_k)}{k} = 0 \ , \tag{6}$$

where $\mathbf{Adv}^{\text{mis}-\text{r}}(\mathcal{E}_k; \mathsf{ChA}_k)$ measures MIS-R-security, as defined in [3]. We refer to the latter property as *weak security*. Additionally, the *strong secrecy capacity* $C_s \leq C_w$ is obtained where we restrict the supremum over those schemes achieving MIS-R security, i.e.,

$$\lim_{k\to\infty} \mathbf{Adv}^{\text{mis-r}}(\mathcal{E}_k; \mathsf{ChA}_k) = 0 \ , \tag{7}$$

Wyner [34] provided a full characterization of $C_w$ in the special case where $\mathsf{ChA}_k$ is a degraded version of $\mathsf{ChR}_k$, i.e., such that there exists a transform $T$ with $T \circ \mathsf{ChR}_k = \mathsf{ChA}_k$, where the composition operator $\circ$ is the straightforward generalization of function composition to randomized transforms. Wyner's result was later generalized by Csiszár and Körner [10]. These results were inherently *non-explicit*: That is, existence of secrecy-capacity achieving schemes is proven via the probabilistic method, and the resulting scheme is neither explicitly given, nor it is guaranteed to be efficient. In fact, to date, only a handful of efficient schemes are known. We briefly survey existing constructions.

Syndrome and coset coding. One particular approach, which dates back to Wyner's original paper [34], in the setting where $\mathsf{ChR}$ is noiseless is a technique known as *syndrome coding*: Given a message $M \in \{0,1\}^m$ and a matrix $H \in \{0,1\}^{m\times k}$, Alice samples a random preimage $R \in \{0,1\}^k$ such that $H \cdot R = M$, and sends $R$ to Bob. Wyner proved that there exists a good choice of the matrix $H$ yielding weak security. This analysis was further improved and extended to the case where $\mathsf{ChR}$ is noisy (and $H$ is applied to a codeword) by Cohen and Zémor [8, 9]. However, all of these schemes only achieve weak security. It is fair to mention that from a construction standpoint, syndrome coding bears some similitude with the extractor-inversion approach introduced Section 5 which we follow, specifically when the given extractor is the two-universal function based on matrix-vector multiplication (which can be shown to be efficiently invertible), even though this approach was not taken by these works, as they did not consider seeded encryption as a goal. Moreover, we stress that no existing proof implies MIS-R-security of these schemes, let alone DS-security.

An alternative way to look at syndrome coding is as a special instance of a more general approach: One takes a (typically linear) code $\mathsf{E} : \{0,1\}^k \to \{0,1\}^n$ which is good for the channel $\mathsf{ChR}$, and then, for a given message set $\{0,1\}^m$, partitions the code $\mathcal{C} = \{\, \mathsf{E}(x) \,:\, x \in \{0,1\}^k \,\}$ in $2^m$ sets as $\mathcal{C} = \bigcup_{M\in\{0,1\}^m} \mathcal{C}_M$. Encryption of $M$ proceeds by selecting a random element of $\mathcal{C}_M$. Usually, one lets $\mathcal{C}_{0^m}$ be a linear subspace of $\mathcal{C}$, called the *inner code* ($\mathcal{C}$ is the *outer code*), and the sets $\mathcal{C}_M$ are the *cosets* in $\mathcal{C}/\mathcal{C}_{0^m}$. Further instantiations of this approach have been considered in [32, 30], but only explicit schemes for a noiseless $\mathsf{ChR}$ and a binary erasure channel $\mathsf{ChA}$ have been obtained, and also only for MIS-R security.

Polar codes. A novel approach has been recently proposed by Mahdavifar and Vardy [25] and by Hof and Shamai [17] (similar ideas also appeared in [20, 1]). They show that polar codes [2] can be used to directly build encryption schemes for the wiretap setting with binary-input symmetric channels. However, these schemes only provide weak security. The full version [24] of [25] provides a variant of the scheme achieving MIS-R-security, which can also be shown to achieve MIS-security (and hence DS-security) as an application of the techniques from [3]; yet, the scheme is only a proof-of-concept, as it is in particular not known how to decrypt its ciphertexts, not even inefficiently. Also note that only recently a first solution to the question of efficiently generating polar codes has appeared [31], which remains an open research direction, and hence relying on this specific code family may be somewhat problematic. Our solution, in contrast, works for arbitrary codes.

Wiretap channel II. Ozarow and Wyner [28] also considered an alternative to the above wiretap setting (called the *wiretap channel II*) where $\mathsf{ChR}$ is noiseless, but at the same time, Eve can learn a fraction $\delta$ of the bits sent over $\mathsf{ChR}$, and does not learn anything about the remaining $(1-\delta)$ fraction. Solutions were presented relying on error-correcting codes [28, 33]. Also, the notable work of [7] noted the such protocols with good parameters can be built from primitives such as deterministic randomness extractors for symbol-fixing sources with efficient inversion [19], as well as from $k$-wise independent functions [21] and related tools from exposure-resilient cryptography, such as all-or-nothing transforms [6, 13].

# B Proof of Lemma 5.2

**Proof:** Note that $\mathsf{Ext}$ is two-universal, as for all distinct $X, X' \in \{0,1\}^k$,

$$\begin{aligned}
&\Pr\left[\, S \leftarrow\!\!\$\ \textsc{Sds} \ :\ \mathsf{Ext}(S,X) = \mathsf{Ext}(S,X') \,\right] \\
&\quad = \Pr\left[\, S \leftarrow\!\!\$\ \textsc{Sds} \ :\ \exists R \in \{0,1\}^{k-m} \setminus \{0^{k-m}\} : S \odot (X \oplus X') = (0;R) \,\right] \\
&\quad \leq \sum_{R \in \{0,1\}^{k-m} \setminus \{0^{k-m}\}} \Pr\left[\, S \leftarrow\!\!\$\ \textsc{Sds} \ :\ S \odot (X \oplus X') = (0;R) \,\right] \leq \frac{2^{k-m}-1}{2^k - 1} = \frac{1}{2^m}\ ,
\end{aligned}$$

since $X \oplus X' \neq 0^k$, and hence there exists at most one $S \in \{0,1\}^k \setminus \{0^k\}$ with $S \odot (X \oplus X') = (0;R)$ (and note that $R \neq 0$); we have additionally used that $\frac{a-1}{b-1} \leq \frac{a}{b}$ for all $a \leq b$. We finally apply the LHL (Lemma 5.1) to conclude the proof. ▮